\documentclass[prl,aps,twocolumn,showpacs,floatfix,superscriptaddress]{revtex4-1}
\usepackage{graphicx}
\usepackage{subfigure}
\usepackage{bm}
\usepackage{amsmath}
\usepackage{tabularx}
\usepackage{wasysym}
\usepackage{array}
\usepackage{multirow}
\usepackage[version=3]{mhchem}
\usepackage{float}
\usepackage{color}
\usepackage{hyperref}
\usepackage{lipsum}
\usepackage{ulem}
\makeatletter
\newcommand{\colorcaption}[2][]{%
  \begingroup%
  \renewcommand{\@caption@fignum@sep}{ (color online). }%
  \caption[#1]{#2}%
  \endgroup%
}
\makeatother

\begin{document}
\title{Distinct Reconstruction Patterns and Spin-Resolved Electronic States along the Zigzag Edges of Transition Metal Dichalcogenides}

\author{Ping Cui}
\affiliation{International Center for Quantum Design of Functional Materials (ICQD), Hefei National Laboratory for Physical Sciences at the Microscale, and Synergetic Innovation Center of Quantum Information and Quantum Physics, University of Science and Technology of China, Hefei, Anhui 230026, China}

\author{Jin-Ho Choi}
\affiliation{International Center for Quantum Design of Functional Materials (ICQD), Hefei National Laboratory for Physical Sciences at the Microscale, and Synergetic Innovation Center of Quantum Information and Quantum Physics, University of Science and Technology of China, Hefei, Anhui 230026, China}
\affiliation{Department of Energy and Materials Engineering and Advanced Energy and Electronic Materials Research Center, Dongguk University-Seoul, Seoul 100-175, Republic of Korea}

\author{Wei Chen}
\affiliation{International Center for Quantum Design of Functional Materials (ICQD), Hefei National Laboratory for Physical Sciences at the Microscale, and Synergetic Innovation Center of Quantum Information and Quantum Physics, University of Science and Technology of China, Hefei, Anhui 230026, China}
\affiliation{Department of Physics and School of Engineering and Applied Sciences, Harvard University, Cambridge, Massachusetts 02138, USA}

\author{Jiang Zeng}
\affiliation{International Center for Quantum Design of Functional Materials (ICQD), Hefei National Laboratory for Physical Sciences at the Microscale, and Synergetic Innovation Center of Quantum Information and Quantum Physics, University of Science and Technology of China, Hefei, Anhui 230026, China}

\author{Chendong Zhang}
\affiliation{Department of Physics, The University of Texas at Austin, Austin, Texas 78712, USA}

\author{Chih-Kang Shih}
\affiliation{Department of Physics, The University of Texas at Austin, Austin, Texas 78712, USA}

\author{Zhenyu Li}
\thanks{Corresponding authors:\\zhangzy@ustc.edu.cn\\zyli@ustc.edu.cn}
\affiliation{International Center for Quantum Design of Functional Materials (ICQD), Hefei National Laboratory for Physical Sciences at the Microscale, and Synergetic Innovation Center of Quantum Information and Quantum Physics, University of Science and Technology of China, Hefei, Anhui 230026, China}

\author{Zhenyu Zhang}
\thanks{Corresponding authors:\\zhangzy@ustc.edu.cn\\zyli@ustc.edu.cn}
\affiliation{International Center for Quantum Design of Functional Materials (ICQD), Hefei National Laboratory for Physical Sciences at the Microscale, and Synergetic Innovation Center of Quantum Information and Quantum Physics, University of Science and Technology of China, Hefei, Anhui 230026, China}


\begin{abstract}
Two-dimensional transition metal dichalcogenides represent an emerging class of materials exhibiting various intriguing properties, and integration of such materials for potential device applications will necessarily encounter creation of different boundaries. Using first-principles approaches, here we investigate the structural, electronic, and magnetic properties along two inequivalent zigzag M and X edges of MX$_{2}$ (M=Mo, W; X=S, Se). Along the M edges, we reveal a previously unrecognized but energetically strongly preferred (2x1) reconstruction pattern, which is universally operative for all the MX$_{2}$, characterized by a self-passivation mechanism through place exchanges of the outmost X and M edge atoms. In contrast, the X edges undergo a more moderate (2x1) or (3x1) reconstruction for MoX$_{2}$ or WX$_{2}$, respectively. We further use the prototypical zigzag MoX$_{2}$ nanoribbons to demonstrate that the M and X edges possess distinctly different electronic and magnetic properties, which are discussed for spintronic and catalytic applications.  
\end{abstract}


\maketitle

Two-dimensional (2D) transition metal dichalcogenides (TMDs) represent an emerging class of materials with immense application potentials in optoelectronics~\cite{Wan12699,Gon132053,Jar141102,Kap141128}, catalysis~\cite{Hin055308,Jar07100,Kib12963}, and lubrication~\cite{Mar9310583,Rap051782}. Integration of such 2D materials into actual devices will necessarily invoke the creation of properly tailored boundaries, whose structural properties vitally influence the overall stability and performance of the devices. For this very reason, extensive efforts have been devoted to investigating the structural properties of TMD edges~\cite{Hel00951,Lau0753,Han1110153,Zho132615,Luc153326}. Different edge structures are also naturally accompanied by their characteristic physical properties. Compelling existing examples include the demonstrations of persistent metallic states~\cite{Bol01196803,Li0816739,Bot09325703,Gib156229,Bru159322}, emergent magnetic orders~\cite{Bot09325703,Voj09125416}, and enhanced catalytic activities~\cite{Hin055308,Jar07100} along structurally different edges of MoS$_{2}$. Such properties, in turn, can be exploited for desirable technological developments. 

Among different possibilities of TMD edges, the most commonly encountered types are the zigzag and armchair edges, which also have their close counterparts in graphene~\cite{Jia091701,Gir091705}. Unlike graphene, which has only one type of zigzag edge~\cite{Son06347,Ata113934,Zan13554,Mag14608}, a TMD material with the chemical formula of MX$_{2}$ possesses two types of zigzag edges: metal-terminated $(10\bar{1}0)$ edge (thereafter labeled as M edge) and chalcogen-terminated $(\bar{1}010)$ edge (X edge). Many novel properties associated with the edges are the consequences of the specific type of termination and the exact bonding configurations due to structural reconstruction along the edges, as recently established for the cases of graphene nanoribbons~\cite{Jia091701,Gir091705,Kos08115502,Kos09073401,Kim132723}. For the widely studied zigzag edges of MoS$_{2}$, a variety of structural models have also been proposed, in many cases invoking different stoichiometric ratios along the edges~\cite{Luc153326,Voj09125416,Ray00129,Bol03085410,Lau04510}, or with the inclusion of foreign species such as hydrogen as passivating elements~\cite{Hel00951,Voj09125416,Bol03085410,Lau04510,Cri025659,Tsa141381}. In contrast, likely reconstruction patterns along the TMD zigzag edges that preserve the M:X=1:2 ratio remain to be fully explored. 

In this Letter, we use first-principles calculations within density functional theory (DFT) to explore systematically the likely reconstruction mechanisms along the two different zigzag edges of MX$_{2}$ (M=Mo, W; X=S, Se) that preserve the M:X=1:2 ratio. We find that the differently terminated edges possess distinctly different reconstruction patterns. Along the M edges, an energetically highly favorable place exchange or self-passivation mechanism is \textit{universally} operative for all the four MX$_{2}$ considered, exhibiting substantial inward displacements of all the first-row M atoms and corresponding outward displacements of half of the second-row X atoms. In the ending structure, the more reactive M atoms are all effectively passivated by the X atoms. In contrast, along the X edges, relatively more moderate (2x1) or (3x1) reconstructions are predicted for MoX$_{2}$ or WX$_{2}$, respectively, characterized by slight local place readjustments of the edge atoms. We further use the prototypical zigzag MoX$_{2}$ nanoribbons to demonstrate that the differently reconstructed M and X edges possess distinctly different electronic and magnetic properties. We also present simulated scanning tunneling microscopy (STM) images along both the reconstructed Mo and X edges, and discuss these results in connection with the standing challenging observations of the ``brim states" along the edges of MoS$_{2}$ nanoclusters on Au(111) substrates ~\cite{Hel00951,Bol01196803,Gib156229,Bru159322}. Collectively, the novel structural and physical properties uncovered along the different edges of the 2D MX$_{2}$ materials are expected to stimulate substantial future research activities in this important area. 

\begin{figure}[tb]
\includegraphics[width=1\columnwidth]{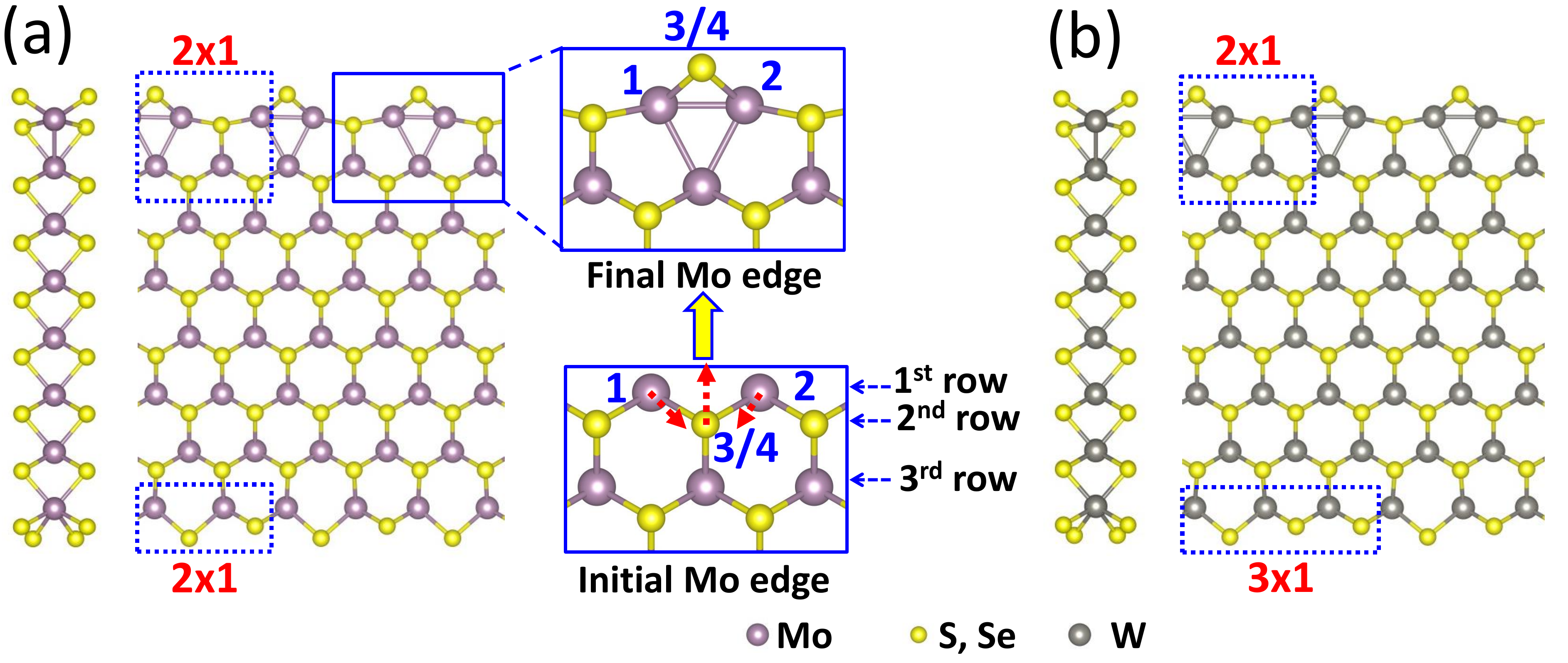}
\colorcaption{ Side view (left panel) and top view (right panel) of optimized atomic structures of 8-zigzag (a) MoX$_{2}$ and (b) WX$_{2}$ (X=S, Se) nanoribbons. The reconstructed edge structures are indicated by the dotted rectangles. The insets in the middle display the place exchange processes involved in the (2x1) reconstruction of the Mo edge.}
\label{fig1}
\end{figure}

The DFT calculations were performed using the projector-augmented wave method~\cite{Blo9417953} implemented in the Vienna \textit{ab initio} simulation package (VASP)~\cite{Kre9611169,Kre991758}, with the Perdew-Burke-Ernzerhof exchange-correlation functional~\cite{Per963865}. More computational details are described in Supplemental Material~\cite{SM}. Along either the M or X edge of a given MX$_{2}$, we first obtain the energies of the unreconstructed, (2x1)-reconstructed, and (3x1)-reconstructed patterns, then identify the most favorable structure. Other possible reconstruction patterns as displayed in Fig. S1~\cite{SM} are found to be energetically less favorable. When specified to the M edge, the (2x1) reconstruction illustrated in Fig. 1 is found to be energetically more favorable for all the four MX$_{2}$ systems. The energy gain for MoS$_{2}$ is 0.261 eV per formula unit along the edge direction relative to the unreconstructed case, and is much larger for the other three MX$_{2}$ systems (Table~\ref{deltaE_edge}). The underlying atomistic mechanism involved is via place exchange, with substantial inward displacements of all the first-row M atoms [Mo 1 and Mo 2 in the insets of Fig. 1(a)] and corresponding outward displacements of half of the second-row X atoms (X 3 and X 4). In the ending structure, the displaced M atoms are dimerized, and each dimer also connects with another M atom belonging to the third row to form a planar M$_{3}$ trimer, with the edge M atoms now all effectively passivated by the displaced X atoms. The corresponding edge-on views are shown in Fig. S2~\cite{SM}. When the M edges are constrained to (3x1) reconstruction patterns, the optimized structures (Fig. S3~\cite{SM}) are metastable, as shown in Table~\ref{deltaE_edge}.

\begin{table}[tb]
\caption{Energy change ($\Delta E$) due to edge reconstruction, defined by $\Delta E = E_{\text{unr}} - E_{\text{rec}}$, where $E_{\text{unr}}$ and $E_{\text{rec}}$ are respectively the total energies of a given system before and after a specific edge reconstruction. The energy change is per formula unit (MX$_{2}$) along the edge direction, given in eV. The values in bold highlight the preferred reconstruction patterns. }
\centering
\begin{tabular}{p{0.125\linewidth}p{0.09\linewidth}<{\centering}p{0.10\linewidth}<{\centering}p{0.09\linewidth}<{\centering}p{0.09\linewidth}<{\centering}p{0.09\linewidth}<{\centering}p{0.09\linewidth}<{\centering}p{0.10\linewidth}<{\centering}p{0.09\linewidth}<{\centering}}
\hline
\hline
  \begin{minipage}{4cm} \vspace{0.3cm}\centering    \vspace{0.3cm}\end{minipage}  & \multicolumn{2}{c}{MoS$_{2}$}  & \multicolumn{2}{c}{MoSe$_{2}$}  & \multicolumn{2}{c}{WS$_{2}$} & \multicolumn{2}{c}{WSe$_{2}$} \\
\cline{2-9}
\begin{minipage}{4cm} \vspace{0.3cm}\centering    \vspace{0.3cm}\end{minipage} & 2x1 & 3x1 & 2x1 & 3x1 & 2x1 & 3x1 & 2x1 & 3x1\\
 \hline
 \begin{minipage}{4cm} \vspace{0.3cm}\centering  \vspace{0.3cm}\end{minipage}  M edge & \textbf{0.261} & -0.014  & \textbf{0.823} & 0.515 & \textbf{0.807} & 0.531 & \textbf{1.012} & 0.690\\
 \hline
 \begin{minipage}{4cm} \vspace{0.3cm}\centering  \vspace{0.3cm}\end{minipage}  X edge  & \textbf{0.320} & 0.259 & \textbf{0.158} & 0.115 & 0.120 & \textbf{0.172} & 0.009 & \textbf{0.046}\\
\hline
\hline
\label{deltaE_edge}
\end{tabular}
\end{table}

For the X edges, the most stable structures of MoX$_{2}$ are (2x1)-reconstructed, with periodically alternating Mo-X bond lengths [Fig. 1(a)]. The short and long Mo-S (Mo-Se) bond lengths are 2.36 (2.50) and 2.46 (2.57) {\text {\AA}}, respectively, with the corresponding edge-on views shown in Fig. S2(a)~\cite{SM}. The second-row Mo atoms are also dimerized, with the distance of 3.02 (3.12) {\text {\AA}} shortened from the bulk-terminated values of 3.18 (3.32) {\text {\AA}} for MoS$_{2}$ (MoSe$_{2}$). The corresponding energy gain is 0.320 (0.158) eV per formula unit along the edge for MoS$_{2}$ (MoSe$_{2}$). The underlying driving force for the distinctly different (2x1) reconstructions along both the Mo and X edges can be uniquely attributed to the same dimerization trend of the quasi-1D Mo atoms along the Mo and X edges. 

For the X edges of WX$_{2}$, the (3x1) reconstructions are found to be most stable, with 2/3 of the W-X bonds shortened and the other 1/3 elongated [Fig. 1(b) and Fig. S4~\cite{SM}]. The second-row W atoms are also linearly trimerized, which may provide the underlying driving force for the overall (3x1) reconstructions. The distance between two neighboring W atoms in a linear trimer is 3.06 (3.19) {\text {\AA}} and the gaps between two neighboring W trimers are 3.43 (3.58) {\text {\AA}} for WS$_{2}$ (WSe$_{2}$).

The contrasting reconstruction geometries at the two zigzag edges of MX$_{2}$ are expected to also result in different electronic and magnetic properties, as exemplified below using the prototypical systems of MoX$_{2}$. We start from the spin-unpolarized band structure of MoS$_{2}$ as a reference, which has two edge states (I and II) crossing the Fermi level [Fig. 2(a)]. Band I is originated from the Mo edge, while band II is from the S edge. When spin polarization is considered, both bands have sizable spin splits (0.3 $\sim$ 0.4 eV). As shown in Fig. 2(b), now the spin-up component of band I (labeled as I-u) becomes fully occupied, while the spin-down component (I-d) is still partially occupied. For band II, the spin-up channel (II-u) is partially occupied, while the spin-down channel (II-d) is totally unoccupied. As a consequence, even though both edges are still metallic, the contributing densities of states around the Fermi level have been significantly reduced, resulting in an overall more stable configuration. The energy gain for the spin-polarized state is 0.071 eV per (2x1) unit cell compared to the spin-unpolarized state. The spatial distributions of the bands at the $\Gamma$ point are displayed in Fig. 2(c). We also note that the I-d and II-u bands are both quite flat, which may provide an important materials-specific platform for exploiting novel physical phenomena involving collectively and/or strongly correlated electrons~\cite{Wu07070401,Liu14077308}.

\begin{figure}[tb]
\includegraphics[width=1\columnwidth]{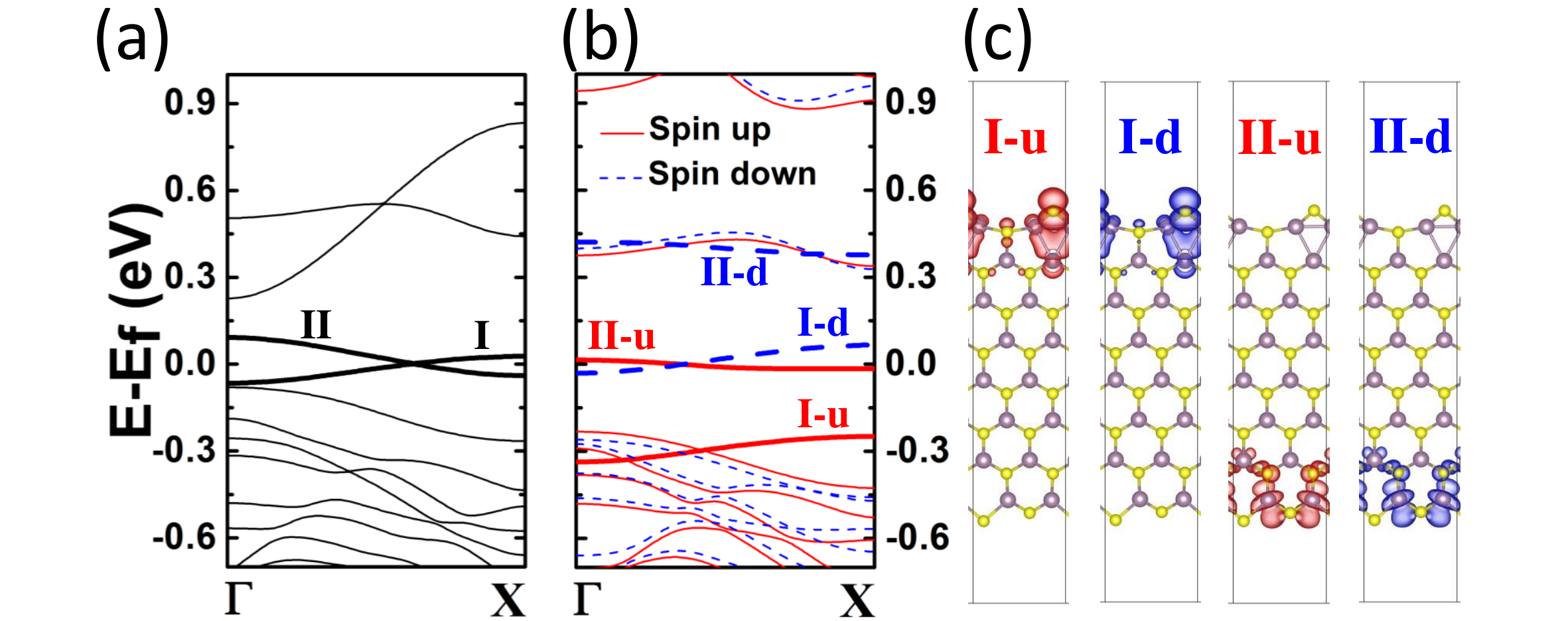}
\colorcaption{(a) Spin-unpolarized and (b) spin-polarized band structures of a (2x1) reconstructed 8-zigzag MoS$_{2}$ nanoribbon. (c) Charge density distributions of the spin-polarized bands crossing the Fermi energy at the $\Gamma$ point. The red and blue colors denote spin-up and spin-down states, respectively. }
\label{fig2}
\end{figure}

As one example, the partially filled flat bands around the Fermi level suggest that ferromagnetism is likely to be stabilized along either reconstructed edge of the MoS$_{2}$ nanoribbon~\cite{Mar886949}. Quantitative confirmations of this expectation are presented in Fig. S5~\cite{SM}, displaying the dominant contributions to the non-zero total magnetic moment by the ferromagnetically coupled constituent atoms along either edge. Furthermore, we find that the total magnetic moment per (2x1) unit cell is 1.38 $\mu_{\text{B}}$ when the two edges are in ferromagnetic (FM) order, and is zero when the two edges are in antiferromagnetic (AFM) order. Since these two edges are well separated, their electronic and magnetic couplings are negligible. Therefore, the AFM band structure shown in Fig. S6(a)~\cite{SM} is essentially the same as the FM band structure shown in Fig. 2(b), but here the two spin channels in band II are mutually exchanged, while the two spin channels in band I stay intact. On a detailed atomistic level, the magnetic moments are located mainly on the outmost S atoms and the inner Mo atoms of the reconstructed Mo$_{3}$ trimers along the Mo edge. This finding is qualitatively different from that shown in a previous systematic study of the magnetic properties at the Mo edges with different Mo:S ratios or with additional passivating hydrogen atoms~\cite{Voj09125416}. In contrast, along the S edge, the magnetic moments are mainly concentrated on the edge Mo atoms, and the outmost S atoms contribute much weaker and AFM-coupled moments. Such magnetic moment  distribution along the S edge is consistent with that presented in the previous study~\cite{Voj09125416}.

\begin{figure}[tb]
\includegraphics[width=1\columnwidth]{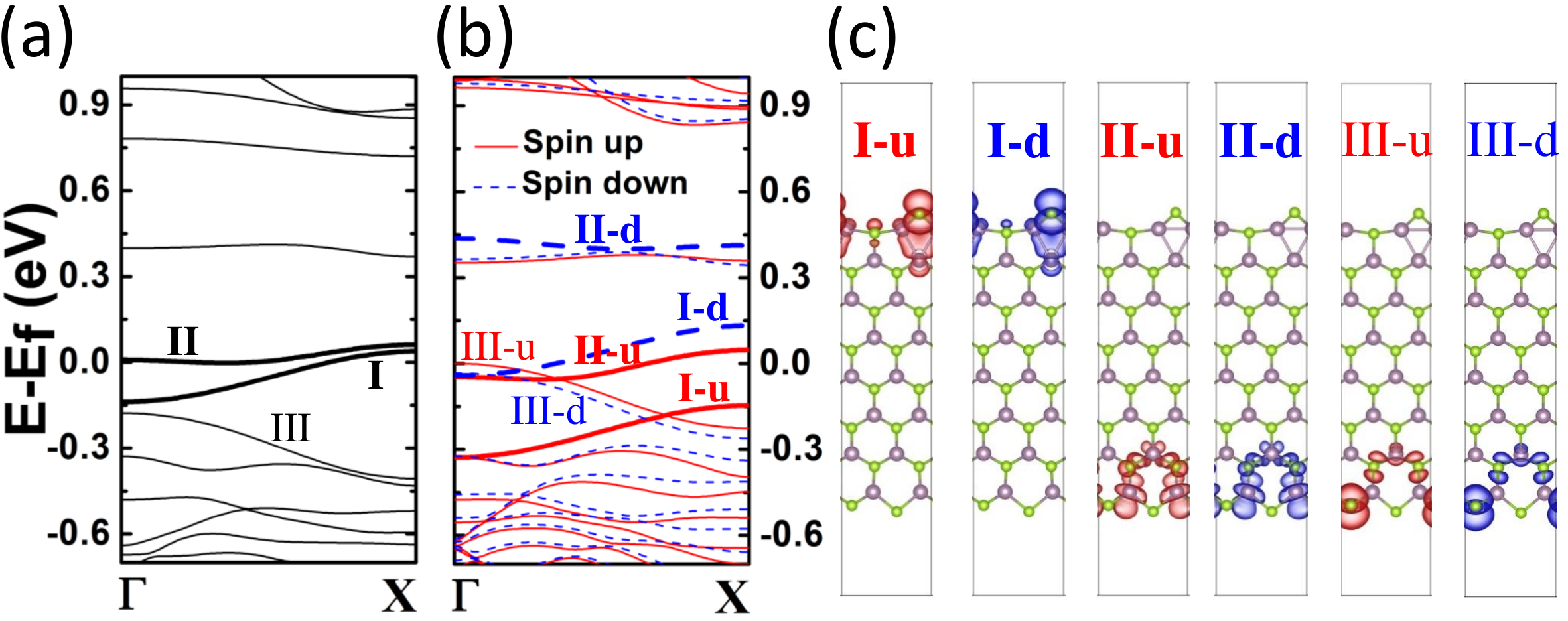}
\colorcaption{(a) Spin-unpolarized and (b) spin-polarized band structures of a (2x1) reconstructed 8-zigzag MoSe$_{2}$ nanoribbon. (c) Charge density distributions of the spin-polarized bands crossing the Fermi energy at the $\Gamma$ point. }
\label{fig3}
\end{figure}

In the spin-unpolarized reference system, the electronic structure of a MoSe$_{2}$ nanoribbon is very similar to that of MoS$_{2}$,  with two edge bands (I and II) crossing the Fermi level [Fig. 3(a)]. When spin polarization is considered, an FM state in the relative magnetic order of the two edges can be obtained, with a total magnetic moment of 1.32 $\mu_{\text{B}}$ per (2x1) unit cell. This FM state is energetically degenerate with the AFM state of the two edges, whose band structure is shown in Fig. S6(b)~\cite{SM}, and both states are 0.069 eV per (2x1) unit cell lower than the spin-unpolarized state. But unlike MoS$_{2}$, for MoSe$_{2}$ another edge state (band III) located farther away from the Fermi level in the spin-unpolarized case moves close to the Fermi level in the spin-polarized case, and splits into bands III-u and III-d [Fig. 3(b)]. These extra bands are derived from the $p_y$/$p_z$ orbitals of the Se atoms contained in the shorter Mo-Se bonds along the Se edge, with band III-u providing another metallic channel. The density distributions of the different spin bands at the $\Gamma$ point are contrasted in Fig. 3(c). 

Recent studies have shown the importance of spin-orbit coupling (SOC) in such TMD systems~\cite{Xia12196802,Xu14343}. As a cross check, we have tested the SOC effects on the edge states of the MoX$_{2}$ ribbons. The results are presented in Fig. S7~\cite{SM}, showing that the spin-resolved electronic structures along the edges stay qualitatively intact in the presence of the SOC. The underlying reason lies in the substantially larger exchange splittings due to the coupling of different spin states along each edge.

For the ribboned structures of such polar materials, electric fields across the ribbon widths may also play an important role in determining their overall stabilities~\cite{Gib156229}. The strengths of such polar fields can be weakened by saturating the M edges with additional X atoms, or by the (2x1) reconstruction along the M edges via the self-passivation mechanism identified here. Indeed, our detailed calculations show that the dipole moment of the 8-zigzag MoS$_{2}$ nanoribbon is reduced from -1.02 e$\cdot${\text {\AA}} per twice the (1x1) unit cell for the unreconstructed structure to 0.30 e$\cdot${\text {\AA}} per (2x1) unit cell for the reconstructed one.

The distinctly different reconstruction patterns and electronic structures should provide an important basis for differentiating such edges once formed experimentally. For this purpose, we have simulated the STM images of the 8-zigzag MoX$_{2}$ nanoribbons, as shown in Fig. 4.  For MoS$_{2}$, at low bias voltages $V_{\text{b}}$ = $\pm$ 0.005 V, the STM images highlight the contributions from the states very close to the Fermi level. We identify a quasi-1D metallic state along either the Mo or S edge [Figs. 4(a) and 4(b)], which corresponds to the spin-polarized band I-d or II-u crossing the Fermi level in Fig. 2(b). At higher bias voltages $V_{\text{b}}$ = $\pm$ 0.1 V, the whole bands crossing the Fermi level are captured and the STM images still have different brightness along the two edges [Figs. 4(c) and 4(d)]. It is interesting to note that, at each of the four bias voltages, the brightest locations along the S edge are not on the outmost atoms, but mainly on the second-row S atoms with dimerized images connected through the bright first-row S atoms, showing the quasi-1D nature of the conducting channel.

\begin{figure}[tb]
\includegraphics[width=1\columnwidth]{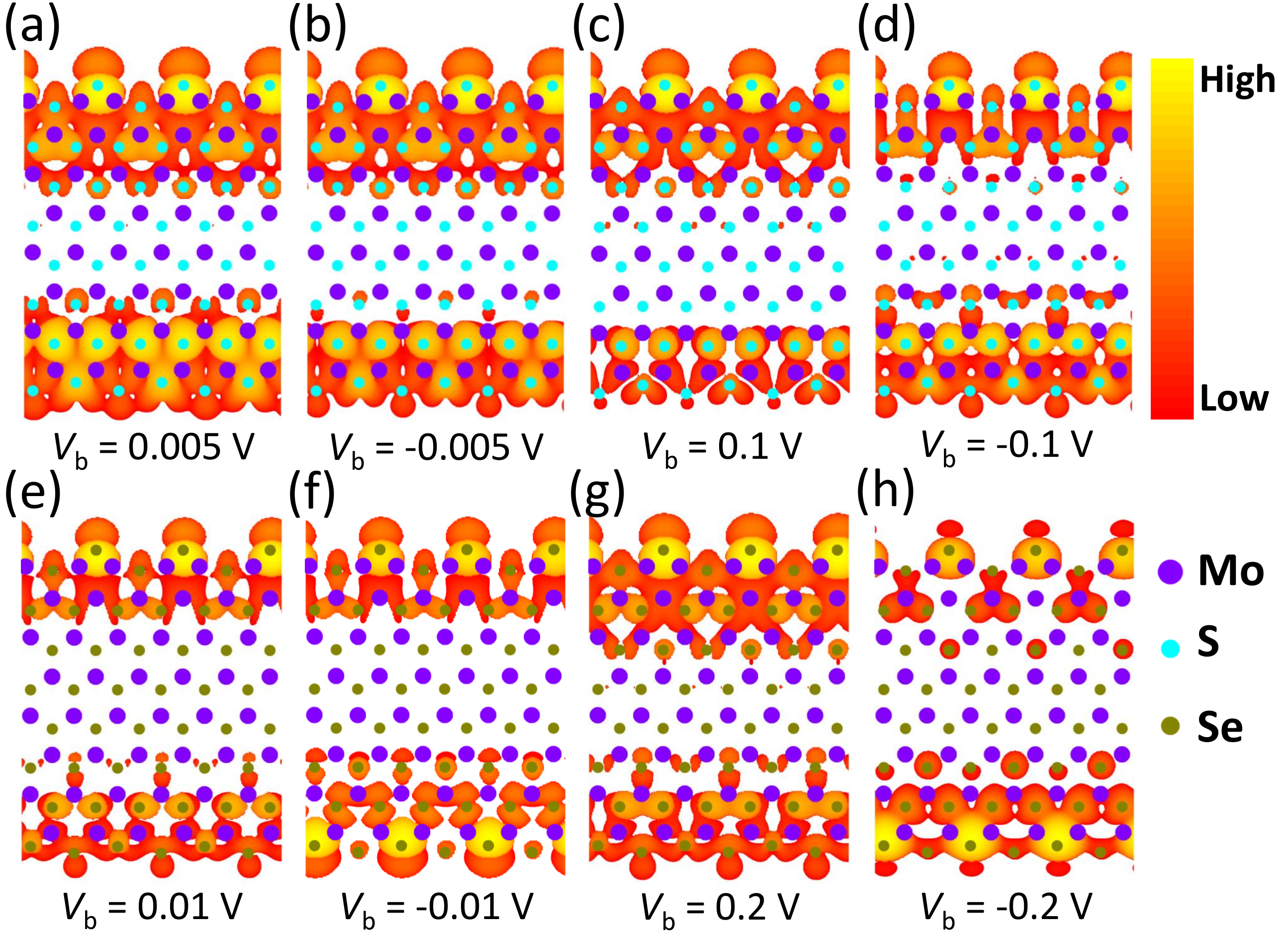}
\colorcaption{Simulated STM images of 8-zigzag (a)-(d) MoS$_{2}$ and (e)-(h) MoSe$_{2}$ nanoribbons at different bias voltages. The MoS$_{2}$ system displays contrasting brim states along the Mo and S edges. }
\label{fig4}
\end{figure}

To a large extent, the electronic structures along the Mo and Se edges of MoSe$_{2}$ are similar to that of MoS$_{2}$. On the other hand, because of the existence of the third band right below the Fermi level [III-u in Fig. 3(b)], the simulated STM images of the MoSe$_{2}$ nanoribbons show different bias dependence from that of the MoS$_{2}$ nanoribbons. In particular, as shown in Figs. 4(e)-4(h), the Se edge displays bright images along the second-row Se atoms at the positive biases, but the first-row Se atoms become brighter at the negative biases.

Next we discuss these predicted edge reconstruction patterns in connection with existing experiments and related theoretical studies. First and foremost, the universal (2x1) reconstruction patterns at the zigzag M edges have no precedence, which call for experimental validations. Secondly, there have been extensive STM studies of zigzag-edged MoS$_{2}$ monolayer nanoislands on Au substrates, revealing striking higher electronic densities centered on the inner rows of the edge atoms; such bright states have been customarily named the brim states~\cite{Hel00951}. To date, the precise origin of the brim states remains to be unambiguously identified~\cite{Bol01196803,Gib156229,Bru159322}. Earlier first-principles studies have primarily focused on associating those brim states along the edges of triangular islands with the Mo edges, but modified by the addition of an S$_{2}$ dimer to each edge Mo atom~\cite{Jar07100,Hel00951,Bol01196803,Voj09125416,Lau04510}. Apparently, the newly identified (2x1) reconstructed structure of the Mo edge with the 1:2 stoichiometric ratio has so far not been considered as a candidate atomic origin for the brim states observed on either the triangular~\cite{Hel00951,Lau04510,Li0716192,Bru159322} or hexagonal islands~\cite{Lau04510}. Here, a closer comparison between the simulated images shown in Figs. 4(a)-4(d) and those shown for the hexagonal islands favors identifying the S edge with the laterally broader brim states (see Fig. 2B in Ref.~\cite{Lau04510}). This tentative identification will naturally leave the other type of edges to be the Mo edges, either self-passivated as proposed here or passivated by extra S$_{2}$ dimers or other atoms. We should also caution that in the experimental situations additional factors can come into play, such as specific shapes of the nanoislands, substrate effects, and possible existence of additional sulfur and/or hydrogen. These factors could in principle cause appreciable deviations between the simulated images under ideal freestanding situations and the experimentally observed images. Given these considerations, we leave definitive identifications of the precise atomic origins of the brim states along different edges of MoS$_{2}$ nanoislands to future efforts, with the central findings of the present study contributing important pieces to the whole picture.

In summary, we have systematically investigated the edge properties of TMDs. Although distinctly different at the two opposite zigzag edges, reconstructions at both edges are energetically favorable. Along the M edges, reconstruction is universally characterized by an intriguing self-passivation mechanism through dramatic place exchanges of the first-row M and second-row X atoms. Reconstructions at the X edges are milder, characterized by varying M-X bond lengths, developing a period of 2 or 3 for MoX$_{2}$ or WX$_{2}$, respectively. These reconstructed edges are also characterized by robust edge states and exotic spin orders, and serve as new candidate atomistic origins for the contrasting brim states observed experimentally along the MoS$_{2}$ nanoislands. Collectively, the novel structural and physical properties uncovered along the different edges of the MX$_{2}$ materials are expected to have important nanoelectronic, spintronic, optical, and catalytic applications.

This work was supported in part by the NSFC (Grant Nos. 61434002, 11204286, 21222304, and 21421063), NKBRPC (Grant No. 2014CB921103), and by the US National Science Foundation (Grant No. EFMA-1542747). The calculations were carried out at the National Supercomputer Center in Tianjin.


\begin{thebibliography}{99}
\bibliographystyle{unsrt}
\bibitem{Wan12699} Q. H. Wang, K. Kalantar-Zadeh, A. Kis, J. N. Coleman, and M. S. Strano, Nat. Nanotechnol.  {\bf 7}, 699 (2012). 
\bibitem{Gon132053} Z. R. Gong, G. B. Liu, H. Y. Yu, D. Xiao, X. D. Cui, X. D. Xu, and W. Yao, Nat. Commun. {\bf 4}, 2053 (2013).
\bibitem{Jar141102} D. Jariwala, V. K. Sangwan, L. J. Lauhon, T. J. Marks, and M. C. Hersam, ACS Nano {\bf 8}, 1102 (2014).
\bibitem{Kap141128} R. Kappera, D. Voiry, S. E. Yalcin, B. Branch, G. Gupta, A. D. Mohite, and M. Chhowalla, Nat. Mater. {\bf 13}, 1128 (2014).
\bibitem{Hin055308} B. Hinnemann, P. G. Moses, J. Bonde, K. P. J{\o}rgensen, J. H. Nielsen, S. Horch, I. Chorkendorff, and J. K. N{\o}skov, J. Am. Chem. Soc. {\bf 127}, 5308 (2005).
\bibitem{Jar07100} T. F. Jaramillo, K. P. J{\o}rgensen, J. Bonde, J. H. Nielsen, S. Horch, and I. Chorkendorff, Science {\bf 317}, 100 (2007). 
\bibitem{Kib12963} J. Kibsgaard, Z. B. Chen, B. N. Reinecke, and T. F. Jaramillo, Nat. Mater. {\bf 11}, 963 (2012). 
\bibitem{Mar9310583} J. M. Martin, C. Donnet, Th. Le Mogne, and Th. Epicier, Phys. Rev. B {\bf 48}, 10583 (1993).
\bibitem{Rap051782} L. Rapoport, N. Fleischer, and R. Tenne, J. Mater. Chem. {\bf 15}, 1782 (2005).
\bibitem{Hel00951} S. Helveg, J. V. Lauritsen, E. L{\ae}gsgaard, I. Stensgaard, J. K. N{\o}rskov, B. S. Clausen, H. Tops{\o}e, and F. Besenbacher, Phys. Rev. Lett. {\bf 84}, 951 (2000).
\bibitem{Lau0753} J. V. Lauritsen, J. Kibsgaard, S. Helveg, H. Tops{\o}e, B. S. Clausen, E. L{\ae}gsgaard, and F. Besenbacher, Nat. Nanotechnol. {\bf 2}, 53 (2007).
\bibitem{Han1110153} L. P. Hansen, Q. M. Ramasse, C. Kisielowski, M. Brorson, E. Johnson, H. Tops{\o}e, and S. Helveg, Angew. Chem. Int. Ed. {\bf 50}, 10153 (2011). 
\bibitem{Zho132615} W. Zhou, X. L. Zou, S. Najmaei, Z. Liu, Y. M. Shi, J. Kong, J. Lou, P. M. Ajayan, B. I. Yakobson, and J. C. Idrobo, Nano Lett. {\bf 13}, 2615 (2013).
\bibitem{Luc153326} M. C. Lucking, J. Bang, H. Terrones, Y.-Y. Sun, and S. B. Zhang, Chem. Mater. {\bf 27}, 3326 (2015).
\bibitem{Bol01196803} M. V. Bollinger, J. V. Lauritsen, K. W. Jacobsen, J. K. N{\o}rskov, S. Helveg, and F. Besenbacher, Phys. Rev. Lett. {\bf 87}, 196803 (2001).
\bibitem{Li0816739} Y. F. Li, Z. Zhou, S. B. Zhang, and Z. F. Chen, J. Am. Chem. Soc. {\bf 130}, 16739 (2008).
\bibitem{Bot09325703} A. R. Botello-M{\'e}ndez, F. L{\'o}pez-Ur{\'i}as, M. Terrones, and H. Terrones, Nanotechnology {\bf 20}, 325703 (2009).
\bibitem{Gib156229} M. Gibertini and N. Marzari, Nano Lett. {\bf 15}, 6229 (2015).
\bibitem{Bru159322} A. Bruix, H. G. F{\"u}chtbauer, A. K. Tuxen, A. S. Walton, M. Andersen, S. Porsgaard, F. Besenbacher, B. Hammer, and J. V. Lauritsen, ACS Nano {\bf 9}, 9322 (2015).
\bibitem{Voj09125416} A. Vojvodic, B. Hinnemann, and J. K. N{\o}rskov, Phys. Rev. B {\bf 80}, 125416 (2009).
\bibitem{Jia091701} X. T. Jia, M. Hofmann, V. Meunier, B. G. Sumpter, J. Campos-Delgado, J. M. Romo-Herrera, H. Son, Y.-P. Hsieh, A. Reina, J. Kong, M. Terrones, and M. S. Dresselhaus, Science {\bf 323}, 1701 (2009).
\bibitem{Gir091705} {\c{C}}. {\"O}. Girit, J. C. Meyer, R. Erni, M. D. Rossell, C. Kisielowski, L. Yang, C.-H. Park, M. F. Crommie, M. L. Cohen, S. G. Louie, and A. Zettl, Science {\bf 323}, 1705 (2009).
\bibitem{Son06347} Y.-W. Son, M. L. Cohen, and S. G. Louie, Nature {\bf 444}, 347 (2006).
\bibitem{Ata113934} C. Ataca, H. {\c{S}}ahin, E. Akt{\"u}rk, and S. Ciraci, J. Phys. Chem. C {\bf 115}, 3934 (2011).
\bibitem{Zan13554} A. M. van der Zande, P. Y. Huang, D. A. Chenet, T. C. Berkelbach, Y. M. You, G.-H. Lee, T. F. Heinz, D. R. Reichman, D. A. Muller, and J. C. Hone, Nat. Mater. {\bf 12}, 554 (2013).
\bibitem{Mag14608} G. Z. Magda, X. Z. Jin, I. Hagym{\'a}si, P. Vancs{\'o}, Z. Osv{\'a}th, P. Nemes-Incze, C. Y. Hwang, L. P. Bir{\'o}, and L. Tapaszt{\'o}, Nature {\bf 514}, 608 (2014).
\bibitem {Kos08115502} P. Koskinen, S. Malola, and H. H{\"a}kkinen, Phys. Rev. Lett. {\bf 101}, 115502 (2008).
\bibitem{Kos09073401} P. Koskinen, S. Malola, and H. H{\"a}kkinen, Phys. Rev. B {\bf 80}, 073401 (2009).
\bibitem{Kim132723} K. Kim, S. Coh, C. Kisielowski, M. F. Crommie, S. G. Louie, M. L. Cohen, and A. Zettl, Nat. Commun. {\bf 4}, 2723 
(2013).
\bibitem{Ray00129} P. Raybaud, J. Hafner, G. Kresse, S. Kasztelan, and H. Toulhoat, J. Catal. {\bf 189}, 129 (2000). 
\bibitem{Bol03085410} M. V. Bollinger, K. W. Jacobsen, and J. K. N{\o}rskov, Phys. Rev. B {\bf 67}, 085410 (2003).
\bibitem{Lau04510} J. V. Lauritsen, M. V. Bollinger, E. L{\ae}gsgaard, K. W. Jacobsen, J. K. N{\o}rskov, B. S. Clausen, H. Tops{\o}e, and F. Besenbacher, J. Catal. {\bf 221}, 510 (2004). 
\bibitem{Cri025659} S. Cristol, J. F. Paul, E. Payen, D. Bougeard, S. Cl{\'e}mendot and F. Hutschka, J. Phys. Chem. B {\bf 106}, 5659 (2002). 
\bibitem{Tsa141381} C. Tsai, F. Abild-Pedersen, and J. K. N{\o}rskov, Nano Lett. {\bf 14}, 1381 (2014).
\bibitem{Blo9417953} P. E. Bl{\"{o}}chl, Phys. Rev. B {\bf 50}, 17953 (1994).
\bibitem{Kre9611169} G. Kresse and J.  Furthm{\"{u}}ller, Phys. Rev. B {\bf 54}, 11169 (1996).
\bibitem{Kre991758} G. Kresse and  D. Joubert, Phys. Rev. B {\bf 59}, 1758 (1999).
\bibitem{Per963865} J. P. Perdew, K. Burke, and M. Ernzerhof, Phys. Rev. Lett. {\bf 77}, 3865 (1996).
\bibitem{Ter831998} J. Tersoff and D. R. Hamann, Phys. Rev. Lett. {\bf 50}, 1998 (1983).
\bibitem{SM} See Supplemental Material at http://link.aps.org/supplemental/XXX for more computational details, edge reconstructed structures, atom-resolved distributions of magnetic moments, and band structures, which includes Refs. [35-39].
\bibitem{Wu07070401} C. J. Wu, D. Bergman, L. Balents, and S. Das Sarma, Phys. Rev. Lett. {\bf 99}, 070401 (2007).
\bibitem{Liu14077308} Z. Liu, F. Liu, and Y. S. Wu, Chin. Phys. B {\bf 23}, 077308 (2014). 
\bibitem{Mar886949} P. M. Marcus and V. L. Moruzzi, Phys. Rev. B {\bf 38}, 6949 (1988). 
\bibitem{Xia12196802} D. Xiao, G. B. Liu, W. X. Feng, X. D. Xu, and W. Yao, Phys. Rev. Lett. {\bf 108}, 196802 (2012).
\bibitem{Xu14343} X. D. Xu, W. Yao, D. Xiao, and T. F. Heinz, Nat. Phys. {\bf 10}, 343 (2014).
\bibitem{Li0716192} T. S. Li and G. Galli, J. Phys. Chem. C {\bf 111}, 16192 (2007).  

\end{thebibliography}
\end{document}